# Interaction patterns of brain activity across space, time and frequency. Part I: methods

Roberto D. Pascual-Marqui[1,2] and Rolando J. Biscay-Lirio[3,4]
1: The KEY Institute for Brain-Mind Research, University Hospital of Psychiatry, Zurich, Switzerland
2: Department of Neuropsychiatry, Kansai Medical University Hospital, Osaka, Japan
3: Institute for Cybernetics, Mathematics, and Physics, Havana, Cuba
4: DEUV-CIMFAV, Facultad de Ciencias, Universidad de Valparaiso, Chile

Corresponding author:
Roberto D. Pascual-Marqui

The KEY Institute for Brain-Mind Research
University Hospital of Psychiatry
Lenggstrasse 31
CH-8032 Zurich
Switzerland
pascualm {at} key.uzh.ch
www.keyinst.uzh.ch/loreta

and:

Department of Neuropsychiatry
Kansai Medical University Hospital
10-15, Fumizono-cho, Moriguchi
Osaka, 570-8507
Japan
pascualr {at} takii.kmu.ac.jp

## Abstract

We consider exploratory methods for the discovery of cortical functional connectivity. Typically, data for the *i*-th subject ($i=1...N_S$) is represented as $\mathbf{X}_i \in \mathbb{R}^{N_V \times N_T}$, corresponding to brain activity sampled at $N_T$ moments in time from $N_V$ cortical voxels. A widely used method of analysis first concatenates all subjects along the temporal dimension, and then performs an independent component analysis (ICA) for estimating the common cortical patterns of functional connectivity. There exist many other interesting variations of this technique, as reviewed in [Calhoun et al. 2009 Neuroimage 45: S163-172].

We present methods for the more general problem of discovering functional connectivity occurring at all possible time lags. For this purpose, brain activity is viewed as a function of space and time, which allows the use of the relatively new techniques of functional data analysis [Ramsay & Silverman 2005: Functional data analysis. New York: Springer]. In essence, our method first vectorizes the data from each subject $\mathbf{X}_i^{vec} \in \mathbb{R}^{(N_T N_V) \times 1}$, which constitutes the natural discrete representation of a function of several variables, followed by concatenation of all subjects. The singular value decomposition (SVD), as well as the ICA of this new matrix will reveal spatio-temporal patterns of connectivity. As a further example, in the case of EEG neuroimaging, $\mathbf{X}_i \in \mathbb{R}^{N_V \times N_\Omega}$ may represent spectral density for electric neuronal activity at $N_\Omega$ discrete frequencies from $N_V$ cortical voxels, from the *i*-th EEG epoch. In this case our functional data analysis approach would reveal coupling of brain regions at possibly different frequencies.

## 1. Introduction

For the sake of simplicity, a particular example will be used for explaining the methods. Straightforward generalizations will be considered in a later Section.

Let $\mathbf{X}_i \in \mathbb{R}^{N_V \times N_T}$ denote brain activity for the $i$-th subject ($i=1...N_S$), sampled at $N_T$ moments in time from $N_V$ cortical voxels. Based on such data, it is of interest to find the interactions between different brain regions. This is the topic of "functional connectivity".

Many methods of analysis exist for the study of functional connectivity. Recent reviews are presented in [1] and [2]. Two methods are of particular interest here: one based on the SVD [3], and the other based on group ICA [1].

Let:

Eq. 1 $\quad \mathbf{Y} = \begin{pmatrix} \mathbf{X}_1 & \mathbf{X}_2 & ... & \mathbf{X}_{N_s} \end{pmatrix} \in \mathbb{R}^{N_V \times (N_S N_T)}$

denote the matrix obtained by the temporal concatenation of the subjects. It will be required that the elements of each row have zero mean, i.e. that the concatenated time series at each voxel have zero mean:

Eq. 2 $\quad \mathbf{Y1} = \mathbf{0}$

where $\mathbf{1}$ and $\mathbf{0}$ are vectors of ones and zeros, respectively.

Consider the spatial covariance matrix:

Eq. 3 $\quad \mathbf{C_{YY}} = \frac{1}{N_S N_T} \mathbf{YY}^T \in \mathbb{R}^{N_V \times N_V}$

and its corresponding correlation matrix:

Eq. 4 $\quad \mathbf{R_{YY}} = \left[ diag(\mathbf{C_{YY}}) \right]^{-1/2} \mathbf{C_{YY}} \left[ diag(\mathbf{C_{YY}}) \right]^{-1/2}$

where the "*diag*" operator returns a diagonal matrix by setting all off-diagonal elements to zero.

Then, as demonstrated by Worsley et al [3], the largest normalized eigenvector of $\mathbf{R_{YY}}$, denoted as $\mathbf{\Gamma_Y} \in \mathbb{R}^{N_V \times 1}$, will detect regions of correlated voxels. In practice, this is achieved by thresholding the brain image corresponding to the eigenvector. Those elements with large absolute value will convey information on the correlated brain regions. The method of Worsley et al [3] was recently extended for the detection of senders, hubs, and receivers of cortical information transactions [4].

A commonly used related approach is known as group ICA with temporal concatenation [1], where the matrix $\mathbf{Y}$ in Eq. 1, which must satisfy the condition in Eq. 2, can be factorized as:

Eq. 5 $\quad \mathbf{Y} = \mathbf{A_Y S_Y}$

with $\mathbf{A_Y} \in \mathbb{R}^{N_V \times K}$, $\mathbf{S_Y} \in \mathbb{R}^{K \times (N_S N_T)}$, and $K \leq rank(\mathbf{Y})$ denoting the number of components. This form of factorization is typical with EEG related data, while the factorization for the transposed of $\mathbf{Y}$ is typical of fMRI data (see e.g. [5]). Ideally, the $K$ time series in the matrix $\mathbf{S_Y}$ should be statistically independent in a strict sense, which can be approximately achieved in many different ways (see e.g. [6]). The columns of the matrix $\mathbf{A_Y}$ contain the spatial components, where each one provides information on the correlated brain regions.

Note: The temporal concatenation of data in Eq. 1 is just one possibility. Spatial concatenation is another example. Other data organization schemes are also possible, such as the three dimensional array in tensorial or PARAFAC analyses (see review in e.g. [1]).

These methodologies are of proven value in the discovery of functional connectivity. When they are interpreted from the point of view of functional data analysis [7], new generalizations can be derived, giving detailed temporal information about the nature of the connectivity patterns. The aim of this study is to present a functional data analysis approach to functional connectivity that allows the discovery of brain interactions across space (cortical locations), time, and frequency.

## 2. Functional data analysis perspective

Typically, measures of connectivity are based on the "similarity" between the time series recorded at two different locations. A simple similarity index is, for instance, the cross-correlation coefficient. However, it is nearly impossible to analyze the massive number of similarities when one considers all possible pairs of voxels at all possible time lags.

A solution to this problem can be obtained by considering the basic data as a function of several variables: space (cortical voxels) and time. This is the approach used in functional data analysis [7]. The data from each subject, consisting of brain activity, is now represented as a vector:

**Eq. 6** $\quad \mathbf{X}_i \in \mathbb{R}^{N_V \times N_T} \rightarrow \mathbf{X}_i^{vec} \in \mathbb{R}^{(N_T N_V) \times 1}$

where the "*vec*" operator transforms a matrix into a vector by stacking the columns of the matrix one underneath the other [8]. Thus, the elements of the vector correspond to brain activity values sampled at points in the (space, time) hyperplane.

The new group data matrix is now defined as follows:

**Eq. 7** $\quad \mathbf{Z} = \begin{pmatrix} \mathbf{X}_1^{vec} & \mathbf{X}_2^{vec} & ... & \mathbf{X}_{N_S}^{vec} \end{pmatrix} \in \mathbb{R}^{(N_T N_V) \times N_S}$

This is the basic idea behind functional data analysis, and it may seem deceptively simple, but in fact it is radically different from any other published form of group analysis [1], [9], [10].

## 3. The functional singular value decomposition (fSVD) approach

Here we apply the SVD method described in [3] to the functional data defined in Eq. 7. The first requirement is to center the data to have zero mean value for the elements of each row, such that:

**Eq. 8** $\quad \mathbf{Z}\mathbf{1} = \mathbf{0}$

as in Eq. 2. Next, consider the very high dimensional covariance matrix:

$\mathbf{C}_{\mathbf{ZZ}} = \dfrac{1}{N_S} \mathbf{Z}\mathbf{Z}^T \in \mathbb{R}^{(N_T N_V) \times (N_T N_V)}$

and its corresponding correlation matrix:

**Eq. 9** $\quad \mathbf{R}_{\mathbf{ZZ}} = \left[ diag(\mathbf{C}_{\mathbf{ZZ}}) \right]^{-1/2} \mathbf{C}_{\mathbf{ZZ}} \left[ diag(\mathbf{C}_{\mathbf{ZZ}}) \right]^{-1/2}$

Then, based on the method of Worsley et al [3], the largest normalized eigenvector of $\mathbf{R}_{ZZ}$, denoted as $\mathbf{\Gamma}_Z \in \mathbb{R}^{(N_T N_V) \times 1}$, will detect the time course of the regions of correlated voxels. In practice, this is achieved by thresholding the time varying brain images corresponding to the eigenvector. Those elements with large absolute value will convey information on the time course of the correlated brain regions.

### 3.1. Interpretation example

For instance, after appropriately thresholding $\mathbf{\Gamma}_Z$, if brain region $A$ at an early latency $\tau_A$ has high values, and is followed by high values in a different brain region $B$ at a later latency $\tau_B$, then the interpretation is that brain regions $A$ and $B$ are cross-correlated with the time lag $|\tau_A - \tau_B|$.

Such cross-spatial and cross-temporal connections can be exposed without having to explore nor calculate and analyze explicitly all pairwise cross-correlations.

### 3.2. A practical algorithm

With respect to the practical aspect of computations, note that it is not necessary to perform the SVD on the very high dimensional correlation matrix $\mathbf{R}_{ZZ}$. All that is needed is the largest left eigenvector of the matrix:

**Eq. 10** $\quad \mathbf{U} = \left\{ \left[ diag(\mathbf{C}_{ZZ}) \right]^{-1/2} \mathbf{Z} \right\} \in \mathbb{R}^{(N_T N_V) \times N_S}$

where $\mathbf{Z}$ must satisfy Eq. 8. Typically, $N_S \ll (N_T N_V)$, which allows for a very efficient calculation, using for instance, the iterative power method.

## 4. The functional independent component analysis (fICA) approach

Consider the functional data matrix defined in Eq. 7, satisfying Eq. 8. The functional ICA model is:

**Eq. 11** $\quad \mathbf{Z} = \mathbf{A}_Z \mathbf{S}_Z$

with $\mathbf{A}_Z \in \mathbb{R}^{(N_T N_V) \times K}$, $\mathbf{S}_Z \in \mathbb{R}^{K \times N_S}$, and $K \leq rank(\mathbf{Z})$ denoting the number of components. As above (see Eq. 5), it is required that the $K$ rows in the matrix $\mathbf{S}_Z$ should be statistically independent in a strict sense, which can be approximately achieved in many different ways (see e.g. [6]). Thus, each component, corresponding to a column of the matrix $\mathbf{A}_Z$, conveys information on the time course of the correlated brain regions related to that component. This means that the interpretation explained above in Subsection 3.1 applies for each independent component here.

## 5. Generalizations

The methods presented here can be applied to other brain activity data, especially when using an EEG tomography such as LORETA [11-13].

In one example the basic data $\mathbf{X}_i \in \mathbb{R}^{N_V \times N_\Omega}$ may represent spectral density for electric neuronal activity at $N_\Omega$ discrete frequencies from $N_V$ cortical voxels, from the $i$-th EEG epoch. In this case our functional data analysis approach would reveal coupling of brain regions at possibly different frequencies. For instance, the method may reveal coupling of frontal gamma activity with occipital theta activity.

In an event related potential (ERP) experiment, when analyzing the collection of single trial epochs, a time-varying spectral analysis will produce extremely high dimensional functional data, consisting of spectral density for electric neuronal activity at $N_\Omega$ discrete frequencies from $N_V$ cortical voxels, as a function of time (relative to stimulus onset), for each stimulus (i.e. each epoch). In this case the functional components, either with fSVD or fICA may reveal coupling of different frequencies at different moments in time between different brain regions.